\documentclass[12pt]{article}
\usepackage[utf8]{inputenc}
\usepackage{amsmath}
\numberwithin{equation}{section}
\usepackage{amsfonts}
\usepackage{amssymb}
\usepackage{slashed}
\usepackage{indentfirst}
\usepackage[framemethod=TikZ]{mdframed}
\mdfdefinestyle{MyFrame}{%
    linecolor=blue,
    outerlinewidth=2pt,
    roundcorner=20pt,
    innertopmargin=\baselineskip,
    innerbottommargin=\baselineskip,
    innerrightmargin=20pt,
    innerleftmargin=20pt,
    backgroundcolor=white!50!white }
      
\begin{document}
\title{Boundary terms in Lovelock gravity from dimensionally continued
Chern-Simons forms}
\author{Theo Verwimp{$^{\dagger}$}\\
e-mail: theo.verwimp@telenet.be}
\renewcommand{\today}{Februari 10, 2021}
\maketitle
For a space-time of dimension D with nonempty boundary, it was shown by Müller-Hoissen [Nucl. Phys. B 337, 709 ( 1990)] that every term in the boundary-terms-modified Lovelock action results from the dimensional continuation of the index theorem for the de Rham complex from an even dimension lower than $D$. Here, the surface corrections are expressed in terms of dimensionally continued Chern-Simons forms and the cancellation
of boundary integrals is checked explicitly by independent variations with respect to the $D$-bein form and the connection one-form.\\
\section{Introduction}
\indent The curvature terms in the Einstein-Hilbert action for four-dimensional gravity, and its most natural extension to arbitrary dimensions, the Lovelock action,[1] contain second derivatives of the metric tensor. This implies that the variational principle applied to these gravitational actions not only gives terms that contribute to the field equations, but also yields an integral over a total divergence containing covariant derivatives of the variations of the metric tensor. For a compact domain of integration, this integral is converted into an integral over
the boundary. This boundary integral involves the derivative of variations of the metric tensor normal to the boundary which are not eliminated by the condition $\delta g=0$ at the boundary. Since in gravity we study the dynamics of the metric, it would be unnatural to impose conditions other than $\delta g=0$ on the boundary: the gravitational action must be extremized, keeping only the geometry on the boundary fixed. Therefore, the boundary integral must be cancelled by the variation of a surface term to be included in the original action. The boundary term must be chosen such that for a metric that satisfies
the field equations, the modified action is an extremum under variations of the metric that disappear at the boundary of the domain of integration over which the action is evaluated, but that can have normal derivatives
different from zero.\\

†Former affiliated with: Physics Department; U.I.A., Universiteit Antwerpen Belgium. On retirement from ENGIE Laborelec, Belgium. 

\pagebreak 

The necessity to supplement the gravitational action
with a boundary integral was first revealed in the course
of an investigation into the Hamiltonian formulation of
general relativity.[2] Later, the same boundary term was
derived in the path-integral approach to quantum
gravity.[3]\\
\indent It was shown by Müller-Hoissen[4] that the dimensional continuation of the Gauss-Bonnet-Chern formula for a manifold with nonempty boundary leads to the correct action for Lovelock gravity. It was checked explicitly that the second normal derivatives in the volume integral of  this action are cancelled by the boundary integral.\\
\indent Within the formalism where the second fundamental form is defined as a matrix of one-forms with only normal components on the boundary, the surface corrections can be expressed in terms of dimensionally continued Chern-Simons forms. Myers[5] checked, for a boundary-terms-modified Lovelock action valid in five or six-dimensions, the cancellation of the boundary integrals for independent variations of the $D$-bein and the connection one-form. We will extend Myers work to the Lovelock Lagrangian in arbitrary dimensions.\\
\indent The paper is organized as follows. In section 2, as a
preparation for the extension to higher dimensions, we review the Einstein action with boundary term in first-order formalism. In section 3, we calculate the index theorem for the de Rham complex in arbitrary dimensions by elaborating the Chern-Simons integral appearing in
the surface term. In section 4, we give the expression for
the modified Lovelock action with the boundary terms expressed as dimensionally continued Chern-Simons forms and verify the cancellation of boundary integrals by variation of the modified action.

\section{Boundary terms in Einstein gravity}
Let $S$ be a semi-Riemannian hypersurface of the space-time manifold $(M,g)$ and $N$ a smooth unit normal field determined in a certain  neighbourhood of every $x\in S$. With,
\begin{equation}
\bot^{A}\medskip_{B}=\delta^{A}\medskip_{B}-eN^{A}N_{B}
\end{equation}\\[-9mm]
the unit tensor of projection in the hypersurface $S$[6], where $e\equiv g(N,N)\in\lbrace \pm 1\rbrace $, the extrinsic curvature of $S$ is defined as
\begin{equation}
K_{AB}\equiv -\bot^{C}\medskip_{A}\bot^{D}\medskip_{B}N_{(D;C)}
\end{equation}\\[-10mm]
Its trace is given by,
\begin{equation}
K\equiv K^{A}\medskip_{A}=g^{AB}K_{AB}=-\perp^{B}\medskip_{A}N^{A}\medskip_{;B}=-N^{A}\medskip_{;A}
\end{equation}\\[-12mm]
\indent For a manifold $M$ with nonempty boundary $\partial M$, the action for Einstein gravity takes the form
\begin{equation}
\bar{I}_{E}=\frac{1}{16\pi G}\lbrace\int_{M}RdV-2 \int_{\partial M}KdS\rbrace
\end{equation}\\[+0mm]
\indent Using differential forms, and the notation $\epsilon_{AB}=\frac{1}{2}\epsilon_{ABCD}\theta^{C}\wedge\theta^{D}$, where $\theta^{A}$ is an orthonormal coframe field and $\epsilon_{ABCD}$ the Levi-Civita tensor with $\epsilon_{0123}=+1$, the Einstein action is written as
\begin{equation}
\bar{I}_{E}=\frac{1}{16\pi G}\lbrace\int_{M}\Omega^{AB}\wedge\epsilon_{AB}- \int_{\partial M}i^{\ast}(\Theta^{AB}\wedge\epsilon_{AB}\rbrace
\end{equation}\\[+1mm]
where $i^{\ast}$ means restriction to $\partial M$ (pullback with the canonical injection map $i: \partial M\rightarrow M)$, $\Omega^{AB}$
is the curvature two-form calculated from the Levi-Civita connection one-form $\mu^{AB}=\Gamma^{AB}\medskip_{C}\theta^{C}$ compatible with the metric tensor $g$, and $\Theta^{AB}$ is the second fundamental\\[-2mm] form defined as[5]
\begin{equation}
\Theta_{AB}=\Theta_{ABC}\theta^{C}=e(N_{A}K_{BC}-N_{B}K_{AC})\theta^{C}
\end{equation}\\[-2mm]
One verifies expression (2.5) using $\Theta^{AB}\wedge\epsilon_{AB}=2eKN^{A}\epsilon_{A},\; \epsilon_{A}=(1/3!)\epsilon_{ABCD}\theta^{B}\wedge\theta^{C}\wedge\theta^{C}$, and by expanding the curvature two-form in terms of the Riemann tensor.\\
\indent In the neighbourhood of the boundary we can chose Gaussian normal coordinates $\lbrace x^{i},x^{n}\rbrace$, such that $x^{n}=0$ is the local equation for $\partial M$. The metric in local Gaussian coordinates takes the form
\begin{equation}
ds^{2}=e(dx^{n})^{2}+g_{ij}(x^{k},x^{n})dx^{i}dx^{j}
\end{equation}\\[-5mm]
Now let $\theta^{A}$ be the orthonormal coframe for these coordinates, where $\theta^{n}=dx^{n}=N$. From definition (2.6) we then obtain using (2.2) that $\Theta_{ij}=0$ and $\Theta_{ni}=\Gamma_{n(ij)}\theta^{j}$. However, for these Gaussian normal coordinates we have that $0=d\theta^{n}=-(1/2)C_{AB}\medskip^{n}\theta^{A}\wedge\theta^{B}$, such\\[-2mm] that the structure constants $C_{AB}\medskip^{n}=-C_{BA}\medskip^{n}$ are all zero. Together with the condition \\[-4mm] for zero torsion $C_{AB}\medskip^{C}=\Gamma^{C}\medskip_{AB}-\Gamma^{C}\medskip_{BA}$, this implies\\[-7mm]
\begin{equation}
\Theta^{ni}=\mu^{ni},\quad \Theta^{ij}=0
\end{equation}
\indent Field equations are obtained from (2.5) by independent variations with respect to the four-bein form $\theta^{A}$ and the connection one-form $\mu^{AB}$. Varying the four-bein gives a contribution only from the volume integral and yields Einstein field equations. Varying the volume integral with respect to the connection one-form yield
\begin{multline}
\delta_{\mu}\int_{M}\Omega^{AB}\wedge\epsilon_{AB}=\int_{M}D(\delta\mu^{AB})\wedge\epsilon_{AB}=\int_{M}d(\delta\mu^{AB}\wedge\epsilon_{AB})=\int_{\partial M}i^{\ast}(\delta\mu^{AB}\wedge\epsilon_{AB})\\=\int_{\partial M}i^{\ast}(\delta_{\mu} \Theta^{AB}\wedge\epsilon_{AB})
\end{multline}
which cancels the variation of the surface term. In deriving (2.9) we made use of $D\epsilon_{AB}=0$ for zero torsion[7], and
\begin{equation}
\delta_{\mu}\Theta^{AB}=\delta\mu^{AB}\quad on \quad \partial M
\end{equation}
whereby we assume that the tangential components of the connection are fixed at the boundary but not its normal components.
 
\section{Index theorem for the de Rham complex in arbitrary dimension}
\indent For a $(D=2m)$-dimensional manifold $M$ with boundary $\partial M$, the Euler characteristic $\chi (M)$ can be expressed as the index of the de Rham complex. The Atiyah-Patodi-Singer index theorem then yields the Gauss-Bonnet identity[8]
\begin{equation}
\chi(M)=c\int_{M}H_{m}(\Omega)-c\int_{\partial M}i^{\ast}Q_{m},\quad c=\frac{1}{(2^{D}\pi^{m}m!)}
\end{equation}
with $H_{m}(\Omega)$ the invariant polynomial of degree $m$, given by
\begin{equation}
H_{m}(\Omega)=\epsilon_{A_{1}\cdot\cdot\cdot A_{D}}\Omega^{A_{1}A_{2}}\wedge \cdot\cdot\cdot\wedge \Omega^{A_{D-1}A_{D}}.
\end{equation}
The surface term is the boundary integral of the Chern-Simons form
\begin{equation}
Q_{m}=m\int^{1}_{0}H_{m}(\Theta,\Omega_{1},\cdot\cdot\cdot,\Omega_{t})dt
\end{equation}
derived from the invariant polynomial $H_{m}(\Omega)$, and where
\begin{equation}
\Omega_{t}\medskip^{AB}=d\mu_{t}\medskip^{AB}+\mu_{t}\medskip^{A}\medskip_{C}\wedge\mu_{t}\medskip^{CB}
\end{equation}
is the curvature of a family of connections
\begin{equation}
\mu_{t}\medskip^{AB}=\mu^{AB}-t\Theta^{AB}
\end{equation}
with $t$ varying from $0$ to $1$.\\
\indent For $D=2$, the Chern-Simons form is
\begin{equation}
Q_{1}=\int^{1}_{0}H_{1}(\Theta)dt=\Theta^{AB}\wedge\epsilon_{AB}
\end{equation}
such that
\begin{equation}
\chi(M^{2})=\frac{1}{4\pi}\int_{M}\Omega^{AB}\wedge\epsilon_{AB}-\frac{1}{4\pi}\int_{\partial M}i^{\ast}(\Theta^{AB}\wedge\epsilon_{AB}).
\end{equation}
Therefore, we can identify the modified Einstein action (2.5) as $\bar{I}_{E}=(1/16\pi G)\bar{I}_{1}$ with $\bar{I}_{1}=4\pi \times$(the dimensional continuation to $D=4$ of the Euler index from $D=2$).\\
\indent For $D=4$ one obtains
\begin{equation}
Q_{2}=2\int^{1}_{0}H_{2}(\Theta,\Omega_{t})dt=2\int^{1}_{0}\Theta^{AB}\wedge\Omega_{t}\medskip^{CD}\wedge\epsilon_{ABCD}dt
\end{equation}
and that the index for the de Rham complex is then given by[8]
\begin{multline}
\chi(M^{4})=\dfrac{1}{32\pi^{2}}\int_{M}\Omega^{AB}\wedge\Omega^{CD}\wedge\epsilon_{ABCD}\\-\dfrac{1}{32\pi^{2}}\int_{\partial M}i^{\ast}\left\lbrace 2\Theta^{AB}\wedge \left(\Omega^{CD}-\frac{2}{3}\Theta^{C}\medskip_{E}\wedge \Theta^{ED}\right)\wedge\epsilon_{ABCD}\right\rbrace 
\end{multline}
The dimensional continuation of Eq.(3.9) yields, as shown in Ref.5, the term quadratic in curvature [the $p=2$ term in Eq.(4.1)], in the boundary-terms-modified Lovelock action.\\
\indent To obtain the boundary corrections for all the terms in the Lovelock action, we calculate the Chern-Simons form (3.3) for arbitrary $m$. This means we must elaborate the integral
\begin{equation}
Q_{m}=m\Theta^{A_{1}A_{2}}\wedge\int^{1}_{0}dt\, \Omega_{t}^{A_{3}A_{4}}\wedge \cdot\cdot\cdot\wedge \Omega_{t}^{A_{2m-1}A_{2m}}\epsilon_{A_{1}\cdot\cdot\cdot A_{2m}}
\end{equation}
We have
\begin{equation}
\Omega_{t}^{AB}=\Omega^{AB}-tD\Theta^{AB}+t^{2}\Theta^{A}_{C}\wedge \Theta^{CB}
\end{equation}
with
\begin{equation}
D\Theta^{AB}=d\Theta^{AB}+\mu^{A}\medskip_{C}\wedge\Theta^{CB}+\Theta^{A}\medskip_{C}\wedge\mu^{CB}
\end{equation}
the exterior covariant derivative of the second fundamental form. With (2.8) it follows that
\begin{equation}
\Omega_{t}\medskip^{ij}\medskip_{\mid \partial M}=\Omega^{ij}+(t^{2}-2t)\Theta^{i}\medskip_{n}\wedge\Theta^{nj} ,
\end{equation}
\begin{equation}
\Omega_{t}\medskip^{nj}\medskip_{\mid\partial M}=(1-t)\Omega^{nj}\medskip_{\mid\partial M}
\end{equation}
where
\begin{equation}
\Omega^{nj}\medskip_{\mid\partial M}=(d\mu^{nj}+\mu^{n}\medskip_{i}\wedge\mu^{ij})_{\mid\partial M}=d\Theta^{nj}+\Theta^{n}\medskip_{i}\wedge\mu^{ij} .
\end{equation}
Since $\Theta^{AB}$ has one normal index at the boundary [see Eq.(2.8)], we find using (3.13) that modulo terms which disappear at the boundary (noted by $\approx$):
\begin{multline}
Q_{m}\approx m \Theta^{A_{1}A_{2}}\wedge \int^{1}_{0} dt[\Omega^{A_{3}A_{4}}+(t^{2}-2t)\Theta^{A_{3}}\medskip_{C}\wedge \Theta^{CA_{4}}]\wedge \cdot\cdot\cdot\\ \wedge[\Omega^{A_{2m-1}A_{2m}}+(t^{2}-2t)\Theta^{A_{2m-1}}\medskip_{C}\wedge \Theta^{CA_{2m}}]\epsilon_{A_{1}\cdot\cdot\cdot A_{2m}}. 
\end{multline}
By the summation over the indices of the Levi-Civita tensor, the binomial formula of Newton applies [the indices of $\Omega^{AB}$ and $\Theta^{C}\medskip_{E}\wedge\Theta^{ED}$ can be changed mutually in the                    \\[-2mm] product of these two two-forms, and the product of the $(m-1)$ factors within the integral sign is commutative]. In this way, one finds
\begin{equation}
Q_{m}\approx\sum_{k=0}^{m-1}b_{k,m}\Phi_{(k,m)}
\end{equation}
with
\begin{multline}
\Phi_{(k,m)}=\Theta^{A_{1}A_{2}}\wedge\Omega^{A_{3}A_{4}}\wedge\cdot\cdot\cdot\wedge\Omega^{A_{2k+1}A_{2k+2}}\wedge\Theta^{A_{2k+3}}\medskip_{B_{1}}\wedge\Theta^{B_{1}A_{2k+4}}\wedge \\ \cdot\cdot\cdot\wedge\Theta^{B_{m-k-1}A_{2m}}\,\epsilon_{A_{1}\cdot\cdot\cdot A_{2m}}
\end{multline}
\begin{equation}
b_{k,m}=\frac{m!}{k!(m-k-1)!}\int_{0}^{1}dt(t^{2}-2t)^{m-k-1}=\frac{(-1)^{m-k+1}m!2^{m-k-1}}{k!\prod_{l=0}^{m-k-1}(2l+1)}\quad\quad\quad\quad\quad
\end{equation}
The index theorem (3.1) for the de Rham complex is then
\begin{equation}
\chi(M^{2m})=c\int_{M}H_{m}(\Omega)-c\int_{\partial M}\sum_{k=0}^{m-1}b_{k,m}i^{\ast}\Phi_{(k,m)}
\end{equation}
After a suitable renaming of the indices in (3.20), the expression can be compared with the result given, for example, in Ref. 9.

\section{Lovelock action with boundary terms}
With the index theorem for the de Rham complex given by (3.20), one obtains through dimensional continuation the Lovelock action modified with boundary terms in dimension $D$,
\begin{equation}
\bar{\mathbb{I}}_{L}=\sum_{p=0}^{[D/2]}a_{p}\bar{\mathbb{I}}_{p}
\end{equation}
where $a_{p}$ are constants, $[D/2]$ denotes the largest integer $\leqslant D/2$, and
\begin{align}
 &\bar{\mathbb{I}}_{p}=\int_{M}\mathbb{R}^{(p)}-\int_{\partial M}i^{\ast}Q_{p},\quad\quad\quad\quad\quad\quad\quad\quad\quad\quad\quad\quad\quad\quad\quad\quad\qquad\qquad\qquad\qquad \\[+2mm]
 &\mathbb{R}^{p}=\Omega^{A_{1}A_{2}}\wedge\cdot\cdot\cdot\wedge\Omega^{A_{2p-1}A_{2p}}\wedge\epsilon_{A_{1}\cdot\cdot\cdot A_{2p}},\\
 &Q_{p}=\sum_{k=0}^{p-1}b_{(k,p)}\Phi_{(k,p)},\\
 &b_{k,p}=(-1)^{p-k+1}2^{p-k-1}p!/[k!1.3.\cdot\cdot\cdot .(2p-2k-1)],
 \end{align}
\begin{multline}
\Phi_{(k,p)}=\Theta^{A_{1}A_{2}}\wedge\Omega^{A_{3}A_{4}}\wedge\cdot\cdot\cdot\wedge\Omega^{A_{2k+1}A_{2k+2}}\wedge\Theta^{A_{2k+3}}\medskip_{B_{1}}\wedge\Theta^{B_{1}A_{2k+4}}\wedge \\ \cdot\cdot\cdot\wedge\Theta^{B_{p-k-1}A_{2p}}\,\epsilon_{A_{1}\cdot\cdot\cdot A_{2p}},
\end{multline}
and finally
\begin{equation}
\epsilon_{A_{1}\cdot\cdot\cdot A_{p}}\equiv[1/(D-k)!]\epsilon_{A_{1}\cdot\cdot\cdot A_{D}}\theta^{A_{p+1}}\wedge\cdot\cdot\cdot\wedge\theta^{A_{D}}.
\end{equation}\\
In (4.1), every $\bar{\mathbb{I}}_{p}$ equals $(1/c)\times$(the dimensional continuation to dimension $D$ of the Euler index of a $2p$-dimensional manifold with boundary).\\
\indent Varying (4.1) with respect to the $D$-bein $\theta$ gives only a contribution from the volume integral and yields the $D$-dimensional field equations. We then only have to show that every $\bar{\mathbb{I}}_{p}$ disappears if one varies the connection one-form $\mu^{AB}$, this is, $\delta_{\mu}\bar{\mathbb{I}}_{p}=0$ for every $p$.\\
\indent From the Bianchi identity, $D\Omega=0$, we obtain
\begin{multline}
d\left\lbrace (\delta\mu^{A_{1}A_{2}})\wedge\Omega^{A_{3}A_{4}}\wedge\cdot\cdot\cdot\wedge\Omega^{A_{2p-1}A_{2p}}\wedge\epsilon_{A_{1}\cdot\cdot\cdot A_{2p}}\right\rbrace \\=D(\delta\mu^{A_{1}A_{2}})\wedge\Omega^{A_{3}A_{4}}\wedge\cdot\cdot\cdot\wedge\Omega^{A_{2p-1}A_{2p}}\wedge\epsilon_{A_{1}\cdot\cdot\cdot A_{2p}}\\-(\delta\mu^{A_{1}A_{2}})\wedge\Omega^{A_{3}A_{4}}\wedge\cdot\cdot\cdot\wedge\Omega^{A_{2p-1}A_{2p}}\wedge D\epsilon_{A_{1}\cdot\cdot\cdot A_{2p}}\qquad
\end{multline}
Together with $\Omega=D\mu$, the divergence theorem, $D\epsilon_{A_{1}\cdot\cdot\cdot A_{2p}}$ for zero torsion, and $\delta\Theta^{AB}=\delta\mu^{AB}$ on the boundary, the variation of (4.2) with respect to the connection one-form yields
\begin{multline}
 \delta\bar{\mathbb{I}}_{p}=p\int_{\partial M}i^{\ast}(\delta_{\mu}\Theta^{A_{1}A_{2}}\wedge\Omega^{A_{3}A_{4}\cdot\cdot\cdot A_{2p-1}A_{2p}}\wedge\epsilon_{A_{1}\cdot\cdot\cdot A_{2p}})\\-\int_{\partial M}\sum_{k=0}^{p-2}b_{k,p}i^{\ast}\delta_{\mu}\Phi_{(k,p)}-\int_{\partial M}b_{p-1,p}i^{\ast}\delta_{\mu}\Phi_{(p-1,p)},\qquad
\end{multline}
where we used the notation
\begin{equation}
 \Omega^{A_{1}A_{2}}\wedge\cdot\cdot\cdot\wedge\Omega^{A_{2n-1}A_{2n}}\equiv \Omega^{A_{1}A_{2}\cdot\cdot\cdot A_{2n-1}A_{2n}}.
 \end{equation} 
For $k=p-1$ we find
\begin{equation}
b_{p-1,p}=p,\qquad\qquad\qquad\qquad\qquad\qquad\qquad\qquad\qquad\qquad\qquad\qquad\qquad\qquad\qquad
\end{equation}
\begin{multline}
\delta_{\mu}\Phi_{(p-1,p)}\approx(\delta_{\mu}\Theta^{A_{1}A_{2}})\wedge\Omega^{A_{3}A_{4}\cdot\cdot\cdot A_{2p-1}A_{2p}}\wedge\epsilon_{A_{1}\cdot\cdot\cdot A_{2p}}\\
+2(p-1)\Theta^{A_{1}A_{2}}\wedge\delta\Theta^{A_{3}}\medskip_{C}\wedge\Theta^{CA_{4}}\wedge\Omega^{A_{5}A_{6}\cdot\cdot\cdot A_{2p-1}A_{2p}}\wedge\epsilon_{A_{1}\cdot\cdot\cdot A_{2p}}.
\end{multline}
Here we used $\Omega=d\mu+\mu\wedge\mu$, the fact that $\Theta^{AB}$ and $\delta\mu^{AB}$ have both one normal index, and the relation (2.8) and (2.10) that are valid at the boundary. The first term in (4.12) yields the counterterm for the first term in (4.9). Therefore,
\begin{multline}
\delta\bar{\mathbb{I}}_{p}=-\int_{\partial M}\sum_{k=0}^{p-2}b_{k,p}i^{\ast}\delta_{\mu}\Phi_{(k,p)}\\
-2p(p-1)\int_{\partial M}i^{\ast}(\Theta^{A_{1}A_{2}}\wedge\delta\Theta^{A_{3}}\medskip_{C}\wedge\Theta^{CA_{4}}\wedge\Omega^{A_{5}A_{6}\cdot\cdot\cdot A_{2p-1}A_{2p}}\wedge\epsilon_{A_{1}\cdot\cdot\cdot A_{2p}}).
\end{multline}
For $k=p-2$ we find
\begin{equation}
b_{p-2,p}=-2p(p-1)/3\qquad\qquad\qquad\qquad\qquad\qquad\qquad\qquad\qquad\qquad\qquad\qquad\quad
\end{equation}
\begin{multline}
\delta_{\mu}\Phi_{(p-2,p)}\approx \delta(\Theta^{A_{1}A_{2}}\wedge\Theta^{A_{2p-1}}\medskip_{B_{1}}\wedge\Theta^{B_{1}A_{2p}})
\wedge\Omega^{A_{3}A_{4}\cdot\cdot\cdot A_{2p-3}A_{2p-2}}\wedge\epsilon_{A_{1}\cdot\cdot\cdot A_{2p}}\\
+2(p-2)\Theta^{A_{1}A_{2}}\wedge\delta\Theta^{A_{3}}\medskip_{C}\wedge\Theta^{CA_{4}}\wedge\Omega^{A_{5}A_{6}\cdot\cdot\cdot A_{2p-3}A_{2p-2}}\\
\wedge\Theta^{A_{2p-1}}\medskip_{B_{1}}\wedge\Theta^{B_{1}A_{2p}}\wedge\epsilon_{A_{1}\cdot\cdot\cdot A_{2p}}.
\end{multline}
Since $\Theta^{AB}$ has one normal index, and since

\begin{equation}
\delta(\Theta^{nA_{2}}\wedge\Theta^{A_{3}}\medskip_{n}\wedge\Theta^{nA_{4}})\wedge\epsilon_{nA_{2}\cdot\cdot\cdot A_{2p}}=3\Theta^{nA_{2}}\wedge\delta\Theta^{A_{3}}\medskip_{n}\wedge\Theta^{nA_{4}}\wedge\epsilon_{nA_{2}\cdot\cdot\cdot A_{2p}},
\end{equation}
it follows after suitable renaming the indices in the first term of (4.15), that this first term yields the counterterm for the last term in (4.13). Therefore,

\begin{multline}
\delta\bar{\mathbb{I}}_{p}=-\int_{\partial M}\sum_{k=0}^{p-3}b_{k,p}i^{\ast}\delta_{\mu}\Phi_{(k,p)}
+\frac{4p(p-1)(p-2)}{3}\int_{\partial M}i^{\ast}(\Theta^{A_{1}A_{2}}\wedge\delta\Theta^{A_{3}}\medskip_{C}\wedge\Theta^{CA_{4}}\\ \wedge\Omega^{A_{5}A_{6}\cdot\cdot\cdot A_{2p-3}A_{2p-2}}\wedge\Theta^{A_{2p-1}}\medskip_{B_{1}}\wedge\Theta^{B_{1}A_{2p}}\wedge\epsilon_{A_{1}\cdot\cdot\cdot A_{2p}}).
\end{multline}
After $(n-1)$ iterations of the steps above, we find
\begin{multline}
\delta_{\mu}\bar{\mathbb{I}}_{p}=-\int_{\partial M}\sum_{k=0}^{p-n}b_{k,p}i^{\ast}\delta_{\mu}\Phi_{(k,p)}
+[(-1)^{n+1}2^{n-1}p(p-1)\cdot\cdot\cdot(p-n+1)/1.3. \cdot\cdot\cdot(2n-3)]\\
 \times\int_{\partial M}i^{\ast}(\Theta^{A_{1}A_{2}}\wedge\delta\Theta^{A_{3}}\medskip_{C}\wedge\Theta^{CA_{4}} \wedge\Omega^{A_{5}A_{6}\cdot\cdot\cdot A_{2p-2n+3}A_{2p-2n+4}}\\
 \wedge\Theta^{A_{2p-2n+5}}\medskip_{B_{1}}\wedge\Theta^{B_{1}A_{2p-2n+6}}\wedge\cdot\cdot\cdot\wedge\Theta^{B_{n-2}A_{2p}}\wedge\epsilon_{A_{1}\cdot\cdot\cdot A_{2p}}).
\end{multline}
For $k=p-n$ we have
\begin{equation}
b_{p-n,p}=(-1)^{n+1}2^{n-1}p(p-1)\cdot\cdot\cdot(p-n+1)/1.3.5. \cdot\cdot\cdot(2n-3)(2n-1),\qquad\quad
\end{equation}

\begin{multline}
\delta_{\mu}\Phi_{(p-n,p)}\approx \delta(\Theta^{A_{1}A_{2}}\wedge\Theta^{A_{2p-2n+3}}\medskip_{B_{1}}\wedge\Theta^{B_{1}A_{2p-2n+4}}\wedge\cdot\cdot\cdot\wedge\Theta^{B_{n-1}A_{2p}})\\
\wedge\Omega^{A_{3}A_{4}\cdot\cdot\cdot A_{2p-2n+1}A_{2p-2n+2}}\wedge\epsilon_{A_{1}\cdot\cdot\cdot A_{2p}}
+2(p-n)\Theta^{A_{1}A_{2}}\wedge\delta\Theta^{A_{3}}\medskip_{C}\wedge\Theta^{CA_{4}}\\
\wedge\Omega^{A_{5}A_{6}\cdot\cdot\cdot A_{2p-2n+1}A_{2p-2n+2}}\wedge\Theta^{A_{2p-2n+3}}\medskip_{B_{1}}
\wedge\Theta^{B_{1}A_{2p-2n+4}}\\
\wedge\cdot\cdot\cdot\wedge\Theta^{B_{n-1}A_{2p}}\wedge\epsilon_{A_{1}\cdot\cdot\cdot A_{2p}}.
\end{multline}
Again, using the fact that $\Theta^{AB}$ has one normal index, and with
\begin{multline}
\delta(\Theta^{nA_{2}}\wedge\Theta^{A_{3}}\medskip_{n}\wedge\Theta^{nA_{4}}\wedge\Theta^{A_{2p-2n+5}}\medskip_{n}\wedge\Theta^{nA_{2p-2n+6}}\wedge\cdot\cdot\cdot\wedge\Theta^{nA_{2p}})\wedge\epsilon_{nA_{2}\cdot\cdot\cdot A_{2p}}
\\=(2n-1)(\Theta^{nA_{2}}\wedge\delta\Theta^{A_{3}}\medskip_{n}\wedge\Theta^{nA_{4}}\wedge\Theta^{A_{2p-2n+5}}\medskip_{n}\wedge\Theta^{nA_{2p-2n+6}}\wedge\cdot\cdot\cdot\wedge\Theta^{nA_{2p}})\wedge\epsilon_{nA_{2}\cdot\cdot\cdot A_{2p}},
\end{multline}
it follows that the first term in (4.20) yields the counterterm for the last term in (4.18) such that, after n iterations,
\begin{multline}
\delta_{\mu}\bar{\mathbb{I}}_{p}=-\int_{\partial M}\sum_{k=0}^{p-n-1}b_{k,p}i^{\ast}\delta_{\mu}\Phi_{(k,p)}
+[(-1)^{n+2}2^n p(p-1)\cdot\cdot\cdot(p-n)/1.3. \cdot\cdot\cdot(2n-1)]\\
 \times\int_{\partial M}i^{\ast}(\Theta^{A_{1}A_{2}}\wedge\delta\Theta^{A_{3}}\medskip_{C}\wedge\Theta^{CA_{4}} \wedge\Omega^{A_{5}A_{6}\cdot\cdot\cdot A_{2p-2n+1}A_{2p-2n+2}}\\
 \wedge\Theta^{A_{2p-2n+3}}\medskip_{B_{1}}\wedge\Theta^{B_{1}A_{2p-2n+4}}\wedge\cdot\cdot\cdot\wedge\Theta^{B_{n-1}A_{2p}}\wedge\epsilon_{A_{1}\cdot\cdot\cdot A_{2p}}).
\end{multline}
After $(p-1)$ iterations one finds that

\begin{multline}
\delta_{\mu}\bar{\mathbb{I}}_{p}=\int_{\partial M}-b_{0,p}i^{\ast}\delta_{\mu}\Phi_{(0,p)}
+[(-1)^{p+1}2^{p-1}p!/1.3.5. \cdot\cdot\cdot(2p-3)]\\
\times\int_{\partial M}i^{\ast}(\Theta^{A_{1}A_{2}}\wedge\delta\Theta^{A_{3}}\medskip_{C}\wedge\Theta^{CA_{4}} \wedge\Theta^{A_{5}}\medskip_{B_{1}}\wedge\Theta^{B_{1}A_{6}}\wedge\cdot\cdot\cdot\wedge\Theta^{B_{p-2}A_{2p}}\wedge\epsilon_{A_{1}\cdot\cdot\cdot A_{2p}}),
\end{multline}
where
\begin{equation}
b_{0,p}=(-1)^{p+1}2^{p-1}p!/1.3.5. \cdot\cdot\cdot(2p-1),\qquad\qquad\qquad\qquad\qquad\qquad\qquad\qquad\qquad
\end{equation}
\begin{equation}
\delta_{\mu}\Phi_{(0,p)}\approx\delta(\Theta^{A_{1}A_{2}}\wedge\Theta^{A_{3}}\medskip_{B_{1}}\wedge\Theta^{B_{1}A_{4}} \wedge\Theta^{A_{5}}\medskip_{B_{2}}\wedge\Theta^{B_{2}A_{6}}\wedge\cdot\cdot\cdot\wedge\Theta^{B_{p-1}A_{2p}})\wedge\epsilon_{A_{1}\cdot\cdot\cdot A_{2p}}.
\end{equation}
Since,
\begin{equation}
\delta_{\mu}\Phi_{(0,p)}\approx (2p-1)(\Theta^{nA_{2}}\wedge\delta\Theta^{A_{3}}\medskip_{n}\wedge\Theta^{nA_{4}} \wedge\Theta^{A_{5}}\medskip_{n}\wedge\Theta^{nA_{6}}\wedge\cdot\cdot\cdot\wedge\Theta^{nA_{2p}})\wedge\epsilon_{A_{1}\cdot\cdot\cdot A_{2p}},
\end{equation}
one finally obtains $\delta_{\mu}\bar{\mathbb{I}}_{p}=0$.

\section{Discussion}
Starting from the integral (3.3) of the Chern-Simons form derived from the invariant polynomial (3.2), we calculated in arbitrary dimensions the index theorem for the de Rham complex of a manifold with nonempty
boundary. The dimensional continuation of these index theorems from all even dimensions smaller than the dimension $D$ of the space-time considered, gives the boundary-terms-modified Lovelock action. Within the formalism used here, this means with the second fundamental form defined as the antisymmetric matrix of one-forms (2.6) with only normal components at the boundary, we checked the cancellation of the boundary integrals explicitly for independent variations of the $D$-bein and the connection one-form.

\newpage 
\begin{center}
\textbf{\Large References}
\end{center}
$[1]$ See T. Verwimp, Class. Quantum Grav. 6, 1655 ( 1989) and references
given there.\\
$[2]$ J. W. York, Phys. Rev. Lett. 28, 1082 (1972).\\
$[3]$ G. W. Gibbons and S. W. Hawking, Phys. Rev. D 15, 2752 (1977).\\
$[4]$ F. Müller-Hoissen, Nucl. Phys. B 337, 709 (1990).\\
$[5]$ R. C. Myers, Phys. Rev. D 36, 392 (1987).\\
$[6]$ C. W. Misner, K. S. Thorne, and J. A. Wheeler, \textit{Gravitation} (Freeman, San Francisco, 1973).\\
$[7]$ A. Trautman, Sympos. Math. 12, 139 (1973).\\
$[8]$ T. Eguchi, P. B. Gilkey, and A. J. Hanson, Phys. Rep. 66, 213
(1980).\\
$[9]$ M. Spivak, \textit{A Comprehensive Introduction to Differential Geometry} (Publish or Perish, Berkeley, CA, 1979), Vols. l-5

\end{document}